\documentclass[11pt,dvips]{article}

\usepackage{epsfig,times} 
\usepackage{picinpar}
\usepackage{wrapfig}
\usepackage{floatflt}
\makeatletter
\renewcommand{\section}{\@startsection{section}{1}{0in}
	{0.4\baselineskip}{0.1\baselineskip}{\Large\bf}}
\renewcommand{\subsection}{\@startsection{subsection}{2}{0in}
	{0.25\baselineskip}{-\baselineskip}{\large\bf}}
\renewcommand{\subsubsection}{\@startsection{subsubsection}{3}{0in}
	{0.1\baselineskip}{-\baselineskip}{\normalsize\bf}}
\makeatother
\begin{document}

%
\makeatletter\newcommand{\ps@icrc}{
\renewcommand{\@oddhead}{\slshape{HE.1.3.16}\hfil}}
\makeatother\thispagestyle{icrc}
%
%

\begin{center}
%
{\LARGE \bf Deformed Lorentz Symmetry\\ 
and Ultra-High Energy Cosmic Rays}
\end{center}

\begin{center}
%
%
{\bf L. Gonzalez-Mestres$^{1,2}$}\\
{\it $^{1}$Laboratoire de Physique Corpusculaire, Coll\`ege de France, 75231 Paris Cedex 05, France\\
$^{2}$L.A.P.P., B.P. 110, 74941 Annecy-le-Vieux Cedex, France}
\end{center}

\begin{center}
{\large \bf Abstract\\}
\end{center}
\vspace{-0.5ex}
%
%
Lorentz symmetry violation (LSV) is often discussed using models
of the $TH\epsilon \mu $ type which involve, basically, energy independent
parameters. However, if LSV is generated at the Planck scale or at some
other fundamental length scale, it can naturally preserve Lorentz symmetry
as a low-energy limit (deformed Lorentz symmetry, DLS). Deformed
relativistic kinematics (DRK) would be consistent with special relativity
in the limit $k$ (wave vector) $\rightarrow ~0$ and allow for a deformed
version of general relativity and gravitation. 
We present an updated discussion of the
possible implications of this pattern for cosmic-ray physics at very high
energy. A $\approx ~10^{-6}$ LSV at Planck scale, leading to
a DLS pattern, would potentially be
enough to produce very important observable effects on the properties of
cosmic rays at the $\approx ~10^{20}~eV$ scale (absence of GZK cutoff,
stability of unstable particles, lower interaction rates, kinematical
failure
of the parton model...). We compare
our approach with more recent similar claims made by S. Coleman and S.
Glashow from models of the $TH\epsilon \mu $ type.
%

\vspace{1ex}

%
%
\section{Status of Special Relativity}
\label{relativity.sec}
A basic physics issue underlies the priority debate: who was (were)
the author(s) of the special relativity theory? It clearly turns out that 
historical arguments are biased by physical prejudices and interpretations.  

H. Poincar\'e was the first author to consistently
formulate the relativity principle stating
(Poincar\'e, 1895):
{\it
"Absolute motion of matter, or, to be more precise, the
relative motion of weighable matter and ether, cannot be disclosed. All that
can be done is to reveal the motion of weighable matter with respect to
weighable matter"}.
He further emphasized the deep meaning of this law of Nature 
when he wrote (Poincar\'e, 1901):
{\it "This principle will be confirmed with increasing precision,
as measurements
become more and more accurate"}.

Several authors have emphasized the role of H. Poincar\'e 
in building relativity
and the relevance of his thought
(Logunov, 1995 and 1997; Feynmann, Leighton, \& Sands,    
1964).
In his
June 1905 paper (Poincar\'e, 1905), published before Einsteins's article
(Einstein, 1905) arrived (on June 30) to the editor,
he explicitly wrote the relativistic transformation law for the
charge density and velocity of motion and applied to gravity
the "Lorentz group", assumed to hold for "forces of whatever origin".
But his priority is sometimes
denied on the grounds that {\it "Einstein essentially
announced the failure of all ether-drift experiments past and future as a
foregone conclusion, contrary to Poincar\'e's empirical bias"} (Miller, 1996),
that Poincar\'e did never {\it "disavow the ether"} (Miller, 1996) or that
{\it "Poincar\'e never challenges... the absolute time of newtonian
mechanics... the ether is not only the absolute space of mechanics... but a
dynamical entity"} (Paty, 1996). It is implicitly assumed that A. Einstein
was right in 1905 when {\it "reducing ether to the absolute space of
mechanics"} (Paty, 1996) and that H. Poincar\'e was wrong because {\it "the
ether fits quite nicely into Poincar\'e's view of physical reality: the ether
is real..."} (Miller, 1996). In fact, there is no 
scientific evidence for such an assumption. 

Modern particle physics has
brought back the concept of a non-empty vacuum where free particles propagate:
without such an "ether" where fields can condense, the standard model
of electroweak interactions could not be written and quark confinement could
not be understood. Modern cosmology is not incompatible with
an "absolute local frame" close to that suggested by the study of cosmic
microwave background radiation. Therefore, the "ether" may well turn out
to be a real entity in the {\it XXI-th} century physics and astrophysics. 
Then, the relativity principle would become
a symmetry of physics, a concept whose paternity was attributed to
H. Poincar\'e by R.P. Feynman (as quoted by Logunov, 1995):
{\it "Precisely Poincar\'e proposed investigating what could be done with the
equations without altering their form. It was precisely his idea to pay
attention to the symmetry properties of the laws of physics"}. 

As symmetries in particle physics are in general violated,
Lorentz symmetry may be broken and an absolute local rest frame may
be detectable through experiments performed beyond some critical scale.
Poincar\'e's special relativity (a symmetry applying to
physical processes) could live with this situation, but not Einstein's
approach such as it was formulated in 1905 (an absolute geometry of space-time
that matter cannot escape). But, is Lorentz symmetry broken?
We discuss here two issues: a) the scale
where we may expect Lorentz symmetry to be violated; b) the physical phenomena
and experiments potentially able to uncover Lorentz symmetry violation (LSV).
Previous papers on the subject are (Gonzalez-Mestres, 1998a, 1998b and 1998c) 
and references therein. We have proposed that Lorentz symmetry be a low-energy
limit, broken following a $k^2$-law ($k$ = wave vector) between the low-energy
region and some fundamental energy (length) scale.  

\section{Lorentz Symmetry As a Low-Energy Limit}
\label{deformed.sec}

Low-energy tests of special relativity have confirmed its validity to an
extremely good accuracy, but the situation at very high energy remains more
uncertain.
If Lorentz symmetry violation (LSV) follows
a $E^2$ law
($E$ = energy), similar to the effective gravitational coupling, it can
be $\approx ~1$ at $E~\approx ~10^{21}~eV$ and $\approx ~10^{-26}$ at $E~
\approx ~100~MeV$ (corresponding to the highest momentum scale involved in
nuclear magnetic resonance experiments), in which case it will escape all
existing low-energy bounds (deformed Lorentz symmetry, DLS). 
If LSV is $\approx ~1$ at Planck scale ($E~
\approx ~10^{28}~eV$), and following a similar law, it will be $\approx
~10^{-40}$ at $E~\approx ~100~MeV$ . Our suggestion is not in contradiction
with Einstein's thought such as it became after he had developed general
relativity. In 1921 , A. Einstein wrote in "Geometry and Experiment" 
(Einstein, 1921):
{\it "The interpretation of geometry advocated here cannot be directly applied
to submolecular spaces... it might turn out that such an extrapolation is
just as incorrect as an extension of the concept of temperature to particles
of a solid of molecular dimensions"}.
It is remarkable that special relativity holds at the
attained accelerator energies, but there is no fundamental reason for this to
be the case above Planck scale.

A typical example of patterns violating Lorentz symmetry at very short distance
is provided by nonlocal models where an absolute local rest frame exists and
non-locality in space is introduced through a
fundamental length scale $a$ where new physics 
is expected to occur (Gonzalez-Mestres, 1997a). 
Such models lead to a deformed
relativistic kinematics (DRK) of the form (Gonzalez-Mestres, 1997a and 1997b):
\equation
E~=~~(2\pi )^{-1}~h~c~a^{-1}~e~(k~a)
\endequation
\noindent
where $h$ is the Planck constant, $c$ the speed of light, $k$ the wave vector
and
$[e~(k~a)]^2$ is a convex
function of $(k~a)^2$ obtained from vacuum dynamics.
Such an expression is equivalent to special relativity in the
small $k$ limit. 
Expanding equation (1) for $k~a~\ll ~1$ , we can write 
(Gonzalez-Mestres, 1997a and 1997c):
\begin{eqnarray}
e~(k~a) & \simeq & [(k~a)^2~-~\alpha ~(k~a)^4~ 
+~(2\pi ~a)^2~h^{-2}~m^2~c^2]^{1/2}
\end{eqnarray}
\noindent
$\alpha $ being a model-dependent constant, in the range $0.1~-~0.01$ for
full-strength violation of Lorentz symmetry at the fundamental length scale,
and {\it m} the mass of the particle. For momentum $p~\gg ~mc$ , we get:
\begin{eqnarray}
E & \simeq & p~c~+~m^2~c^3~(2~p)^{-1}~ 
-~p~c~\alpha ~(k~a)^2/2~~~~~
\end{eqnarray}
The "deformation" $\Delta ~E~=~-~p~c~\alpha ~(k~a)^2/2$ in the right-hand
side of (3) implies a Lorentz symmetry violation in the ratio $E~p^{-1}$
varying like $\Gamma ~(k)~\simeq ~\Gamma _0~k^2$ where $\Gamma _0~
~=~-~\alpha ~a^2/2$ . If $c$ is a universal parameter for all
particles, the DRK defined by (1) and (2) preserves Lorentz symmetry 
in the limit $k~\rightarrow ~0$, contrary to the standard
$TH\epsilon \mu $ model (Will, 1993). If $\alpha $ is universal,
LSV does not lead (Gonzalez-Mestres, 1997a, c and e)
to the spontaneous decays predicted in
(Coleman, \& Glashow, 1997 and subsequent papers). 
On more general grounds,
as we also pointed out, the existence of very high-energy cosmic rays
can by no means 
be regarded as an evidence against LSV (Gonzalez-Mestres, 1997d and 1997e). 
The above non-locality may actually be an approximation to
an underlying dynamics involving superluminal particles
(Gonzalez-Mestres, 1996, 1997b, 1997f and 1997g), 
just as electromagnetism looks nonlocal
in the potential approximation to lattice dynamics in solid-state physics:
it would then correspond to the limit $c~c_i^{-1}~\rightarrow~0$
where $c_i$ is the superluminal critical speed. Contrary to the
$TH\epsilon \mu $-type scenario considered
by Coleman and Glashow, where LSV
occurs explicitly in the lagrangian already at $k~=~0$ , our DLS
approach can preserve standard gravitation and general relativity as
low-energy limits. Gravitation can naturally be associated to fluctuations
of the classical parameters (e.g. the parameters of a differential or nonlocal 
equation on classical fields) governing dynamics at the fundamental-length
scale (Gonzalez-Mestres, 1997a). This would be impossible with the 
$TH\epsilon \mu $ approach used in (Coleman, \& Glashow, 1997). More recent
(1998) papers by these authors bring no new result as compared to our
1997 papers and present the same fundamental limitation as their 1997 article. 

Are $c$ and $\alpha $ universal? This may be
the case for all "elementary" particles, i.e.
quarks, leptons, gauge bosons...,
but the situation is less obvious for hadrons, nuclei and heavier objects.
From a naive soliton model (Gonzalez-Mestres, 1997b and 1997f), 
we inferred that: a) $c$ is
expected to be universal up to very small corrections ($\sim 10^{-40}$)
escaping all existing bounds; b)
an
approximate rule can be to take $\alpha $ universal for leptons, gauge bosons
and light hadrons (pions, nucleons...) and assume a $\alpha \propto m^{-2}$
law for nuclei and heavier objects, the nucleon mass setting the scale.
With this rule, DRK introduces no anomaly in the relation between inertial
and gravitational masses at large scale (Gonzalez-Mestres, 1998c).   

\section{Ultra-High Energy Cosmic-Ray Physics}
\label{uhcr.sec}

If Lorentz symmetry is broken at Planck scale or at some other
fundamental length scale, the effects of LSV may be accessible to experiments
well below this
energy: in particular, they can produce detectable phenomena at the highest
observed cosmic ray energies. This is, in particular, due to DRK
(Gonzalez-Mestres 1997a, 1997b, 1997c 1997h and 1998a): at energies above
$E_{trans}~
\approx ~\pi ^{-1/2}~ h^{1/2}~(2~\alpha )^{-1/4}~a^{-1/2}~m^{1/2}~c^{3/2}$,
the very small deformation $\Delta ~E$
dominates over
the mass term $m^2~c^3~(2~p)^{-1}$ in (3) and modifies all
kinematical balances. Because of the negative value of $\Delta ~E$ , it costs
more and more energy, as energy increases above $E_{trans}$,
to split the incoming logitudinal momentum.
With such a LSV pattern, the parton model (in any version), as well as standard
formulae for Lorentz contraction and time dilation, are also expected to fail
above this energy (Gonzalez-Mestres, 1997b and 1997f) which corresponds to $E
~\approx~10^{20}~eV$
for $m$ = proton mass and
$\alpha ~a^2~\approx ~10^{-72}~cm^2$ (f.i. $\alpha
~\approx ~10^{-6}$ and $a$ = Planck
length), and to $E~\approx ~10^{18}~eV$ for
$m$ = pion mass and $\alpha ~a^2~\approx ~10^{-67}~cm^2$
(f.i. $\alpha ~\approx ~0.1$ and $a$ = Planck length).
Assuming that the earth moves slowly with
respect to the absolute rest frame
(the "vacuum rest frame"), these
effects lead to observable phenomena
in future experiments devoted to the highest-energy cosmic rays:

a) For $\alpha ~a^2~>~10^{-72}~cm^2$ , 
assuming universal values of $\alpha $ and $c$ ,
there is no Greisen-Zatsepin-Kuzmin (GZK)
cutoff for the particles under
consideration. Due to the new kinematics, 
interactions with microwave background photons are strongly inhibited or
forbidden, and ultra-high energy cosmic rays (e.g. protons)
from anywhere in the presently observable Universe
can reach the earth (Gonzalez-Mestres, 1997a and 1997c).

b) With the same hypothesis,
unstable particles with at
least two stable particles in the final states
of all their decay channels become stable at very
high energy. Above $E_{trans}$, the lifetimes of all
unstable particles (e.g. the $\pi ^0$ in
cascades) become much longer than predicted
by relativistic kinematics (Gonzalez-Mestres, 1997a, 1997b and 1997c).
Then, for instance,
the neutron or even the $\Delta ^{++}$ can be candidates for the
primaries of the highest-energy cosmic ray
events. If $c$ and $\alpha $ are not exactly universal, 
many different scenarios can happen concerning the stability of 
ultra-high-energy particles
(Gonzalez-Mestres, 1997a, 1997b and 1997c).

c) In astrophysical processes at very
high energy,
similar mechanisms can inhibit radiation under
external forces (e.g. synchrotron-like), GZK-like cutoffs, decays,
photodisintegration of nuclei, momentum loss trough
collisions, production of lower-energy secondaries...
potentially contributing to solve all basic problems
raised by the highest-energy cosmic rays (Gonzalez-Mestres, 1997e).

d) With the same hypothesis, the allowed final-state
phase space of two-body collisions is modified by DRK at very high energy and
can lead to a sharp fall of cross-sections
for incoming cosmic ray energies above
$E_{lim} ~\approx ~(2~\pi )^{-2/3}~(E_T~a^{-2}~ \alpha ^{-1}~h^2~c^2)^{1/3}$,
where $E_T$ is the energy of the target. As a consequence, and with the
previous figures for Lorentz symmetry violation, above some
energy $E_{lim}$ between 10$^{22}$ and $10^{24}$ $eV$ a cosmic
ray will not deposit most of its energy in the atmosphere
and can possibly fake an exotic event with much less energy
(Gonzalez-Mestres, 1997e). Actually, requiring the absence of GZK cutoff above
$\approx ~
10^{20}~eV$ and that cosmic rays with
energies below $\approx ~3.10^{20}~eV$ deposit most of their energy in the
atmosphere, leads in the DRK scenario to the constraint:
$10^{-72}~cm^2~<~\alpha ~a^2~<~
10^{-61}~cm^2$~, equivalent to $10^{-20}~<~\alpha ~<~10^{-9}$ for
$a~\approx 10^{-26}~cm$~. Remarkably enough, assuming full-strength
LSV forces $a$ to be in the range $10^{-36}~cm~<~a~<~
10^{-30}~cm$~. But a $\approx 10^{-6}$ LSV at Planck scale
can still fit the data.

e) Effects a) to d) are obtained using only DRK. If dynamical
anomalies are added (failure, at very small distance scales,
of the parton model and of the
standard Lorentz formulae for length and time...), we can expect
much stronger effects in the cascade development profiles
of cosmic-ray events (Gonzalez-Mestres, 1997b, 1997f and 1998a).
Detailed data analysis in next-generation experiments may therefore uncover 
spectacular new physics and provide a powerful microscope directly
focused on the fundamental length (Planck?) scale.

f) Cosmic superluminal particles would produce atypical events
with very small total momentum, isotropic or involving several
jets 
(Gonzalez-Mestres, 1996, 1997b, 1997d, 1997 and 1998b).

%
%
%
\vspace{1ex}
\begin{center}
{\Large\bf References}
\end{center}
%
Coleman, S., \& Glashow, S., 1997, Phys. Lett. B 405, 249 and subsequent (1998)
papers
in LANL (Los Alamos) electronic archive as well as in the TAUP 97 Proceedings.\\
Einstein, A., 1905, "Zur Elecktrodynamik bewegter K\"{o}rper",
Annalen der Physik 17, 891.\\
Einstein, A., 1921, "Geometrie und Erfahrung",
{\it Preussische Akademie der Wissenchaften,
Sitzungsberichte}, part {\bf I}, p. 123. \\
Feynman, R.P., Leighton, R.B., \& Sands, M., 1964 
"The Feynman Lectures on Physics", Vol. I, Addison-Wesley.\\
Gonzalez-Mestres, L., 1996, paper hep-ph/9610474 of LANL 
electronic archive and references therein.\\
Gonzalez-Mestres, L., 1997a, paper
physics/9704017 of LANL archive.\\
Gonzalez-Mestres, L., 1997b, Proc. of the International
Conference on Relativistic Physics and some of its Applications, Athens June
25-28 1997, Apeiron, p. 319.\\
Gonzalez-Mestres, L., 1997c, Proc. 25th ICRC (Durban, 1997), Vol. 6, p. 113.\\
Gonzalez-Mestres, L., 1997d, papers physics/9702026
and physics/9703020 of LANL archive.\\
Gonzalez-Mestres, L., 1997e, papers physics/9706022 and
physics/9706032 of LANL archive.\\
Gonzalez-Mestres, L., 1997f, paper physics/9708028 of LANL archive.\\
Gonzalez-Mestres, L., 1997g, Proc. 25th ICRC (Durban, 1997), Vol. 6, p. 109.\\
Gonzalez-Mestres, L., 1997h, talk given at the International
Workshop on Topics on Astroparticle and Underground Physics (TAUP 97),
paper physics/9712005 of LANL archive.\\
Gonzalez-Mestres, L., 1998a, talk given at the "Workshop on Observing 
Giant Cosmic Ray Air Showers From $>$ 10$^{20}$ Particles From Space", College
Park, November 1997, AIP
Conference Proceedings 433, p. 148.\\
Gonzalez-Mestres, L., 1998b, same Proceedings, p. 418.\\ 
Gonzalez-Mestres, L., 1998c, Proc. of COSMO 97, Ambleside September 1997,
World Scientific, p. 568.\\
Logunov, A.A., 1995, "On the articles by Henri Poincar\'e on the
dynamics of the electron", Ed. JINR, Dubna.\\
Logunov, A.A., 1997,  
"Relativistic theory of gravity and the
Mach principle", Ed. JINR, Dubna.\\
Miller, A.I. 1996, Why did Poincar\'e not formulate Special
Relativity in 1905?", in "Henri Poincar\'e, Science and Philosophy", Ed.
Akademie Verlag, Berlin, and Albert Blanchard, Paris.\\
Paty, M., 1996, "Poincar\'e et le principe de relativit\'e",
same reference.\\
Poincar\'e, H., 1895, "A propos de la th\'eorie de M. Larmor",
L'Eclairage \'electrique, Vol. 5,  5.\\
Poincar\'e, H., 1901, "Electricit\'e et Optique: La lumi\`ere
et les th\'eories \'electriques", Ed. Gauthier-Villars, Paris.\\
Poincar\'e, H., 1905, Sur la dynamique de l'\'electron", Comptes
Rendus Acad. Sci. Vol. 140,
p. 1504, June 5.\\
Will, C., 1993, "Theory and Experiment in
Gravitational Physics", Cambridge University Press.
\end{document}